\documentclass[conference]{IEEEtran}
\IEEEoverridecommandlockouts
\usepackage{cite}
\usepackage{amsmath,amssymb,amsfonts}
\usepackage{algorithmic}
\usepackage{graphicx}
\usepackage{textcomp}
\usepackage{xcolor}
\def\BibTeX{{\rm B\kern-.05em{\sc i\kern-.025em b}\kern-.08em
    T\kern-.1667em\lower.7ex\hbox{E}\kern-.125emX}}

\usepackage{newpxtext,newpxmath}
\usepackage[hyphens]{url}
\usepackage{hyperref}
\hypersetup{
    colorlinks,
    linkcolor={red!50!black},
    citecolor={blue!60!black},
    urlcolor={blue!80!black}
}
\usepackage{booktabs}
\newcommand{\ra}[1]{\renewcommand{\arraystretch}{1.2}}
\usepackage{enumitem}
\usepackage{lipsum}
\makeatletter
\def\footnoterule{\relax%
  \kern-5pt
  \hbox to \columnwidth{\hfill\vrule width 1.0\columnwidth height 0.4pt\hfill}
  \kern4.6pt}
\makeatother
\usepackage{balance}

\begin{document}

\title{Optimal Bidding Strategy for Maker Auctions}

\author{
    \IEEEauthorblockN{
        Michael Darlin\IEEEauthorrefmark{1}, Nikolaos Papadis\IEEEauthorrefmark{2}, Leandros Tassiulas\IEEEauthorrefmark{2}
    }
    \vspace{0.1cm}
    \IEEEauthorblockA{
        \IEEEauthorrefmark{1}School of Management, and Yale Institute for Network Science, Yale University
    }
    \IEEEauthorblockA{
        \IEEEauthorrefmark{2}Department of Electrical Engineering, and Yale Institute for Network Science, Yale University
    }
    \thanks{The authors thank Florian Ederer for his helpful comments. All errors are our own.}
    \\[-3.5ex]
}

\maketitle
\thispagestyle{plain}
\pagestyle{plain}

\begin{abstract}
The Maker Protocol (``Maker'') is a decentralized finance application that enables collateralized lending. The application uses open-bid, second-price auctions to complete its loan liquidation process. In this paper, we develop a bidding function for these auctions, focusing on the costs incurred to participate in the auctions. We then optimize these costs using parameters from historical auction data, and compare our optimal bidding prices to the historical auction prices. We find that the majority of auctions end at higher prices than our recommended optimal prices, and we propose several theories for these results.
\end{abstract}

\section{Introduction}
Auctions have been used in numerous ways, in both online and offline contexts. A little-studied area has been the use of auctions on public blockchains, and particularly auctions used in the context of decentralized finance (``DeFi'') applications on Ethereum. Because all transactions on the Ethereum blockchain are public, auctions conducted by DeFi applications can be studied in a great level of detail.

This paper examines the auctions of one particular DeFi application, the Maker Protocol (``Maker''). This paper contains the following insights.
\begin{enumerate}
    \item \textbf{Auction overview}: We describe the process by which Maker auctions are executed, as well as the characteristics of the auctions within the framework of formal auction theory.
    \item \textbf{Optimal bidding strategy}: We outline a conceptual model to understand bidder valuations. We then apply the conceptual model to  arrive at a proposed bidding strategy, which focuses on optimizing participation costs.
    \item \textbf{Historical comparison}: We solve for the optimization of participation costs, based on parameters from historical auctions, and use these participation costs to recommend optimal bidding prices. We then compare our recommended bidding prices to actual auction prices, and propose reasons for differences between the proposed strategy and historical results.
\end{enumerate}

The paper is organized in the following manner: Sections~\ref{sec:DeFi} and ~\ref{sec:maker} provide context on DeFi and the Maker Protocol, respectively. Section~\ref{sec:strategy} proposes a conceptual model to understand Maker auctions, and then proposes an optimal bidding strategy. Section~\ref{sec:costs} then defines the costs to participate in Maker auctions. Section~\ref{sec:optimize} describes the process for optimizing participation costs, and then compares the optimized participation costs to historical auctions on the blockchain. Section~\ref{sec:conclusion} provides concluding remarks. Appendix~\ref{section:characteristics} defines Maker auctions within the framework of formal auction theory.

\section{Decentralized finance}
\label{sec:DeFi}
Ethereum, first launched in 2015, is a blockchain network powered by a Proof-of-Work algorithm, with Ether (``ETH'') as its currency. The network's defining feature is its ability to execute smart contracts on the Ethereum Virtual Machine \cite{ethWhitePaper}. The network has attracted a range of potential use-cases, with varying degrees of feasibility. 

One of Ethereum's most visible use-cases has been the enablement of DeFi applications. These applications use Ethereum smart contracts to enable financial transactions, ranging from the relatively simple (lending and borrowing) to the more complex (synthetic asset trading and liquidity pooling) \cite{wsjDefi}. The transactions are commonly performed with stablecoins, which are cryptocurrencies with values intended to be pegged to the US dollar at a 1:1 ratio. Stablecoins are often hosted by the same DeFi applications that enable lending and borrowing \cite{wsjStablecoins}.

A common metric for DeFi usage is Total Value Locked (``TVL''), which measures the amount of currency held in smart contracts used to conduct DeFi transactions. As of December 31, 2018, TVL was measured at \$275M; by July 31, 2020, TVL was measured at \$4.0B\cite{defiPulse}. While TVL has several limitations in capturing the true value of DeFi applications \cite{etoro, consensys.Q2Defi}, its increase implies a general rise in DeFi usage since the beginning of 2019.

Formal research on DeFi applications is relatively scarce, as DeFi is a nascent technology in the only recently-established field of cryptocurrency technology. Relevant research includes i) overviews of DeFi from an economic and legal perspective \cite{defiOverview1, defiOverview2}; ii) analyses of actual or potential exploits for DeFi applications \cite{exploit1, exploit2}; and iii) definitions of mathematical characteristics for DeFi applications \cite{defiMath1, defiMath2}.

\section{Maker Protocol}
\label{sec:maker}
\subsection{Overview}
The Maker Protocol, which was created in 2014 \cite{makerWhitePaper}, is a DeFi application whose primary purpose is to facilitate the creation of the DAI stablecoin. Users can send cryptocurrency (for example, ETH\footnote{The most common collateral used in the Maker Protocol is ETH, or more precisely the ERC-20 compliant version called wrapped ETH (WETH). For simplicity, we hereafter assume that WETH is the collateral being used in the Maker application.}) to a Maker smart contract, which is referred to as the user's ``vault.'' The cryptocurrency deposited in the vault can be used to create DAI, which is recorded as a debt to the user. While the debt remains outstanding, the original currency in the vault is ``locked'' and serves as collateral for the outstanding debt. Maker requires users to maintain a minimum ``collateral ratio'', which is the ratio between the value of locked collateral and the debt. If the collateral ratio falls below a certain threshold,\footnote{As of this writing, the collateral ratio is set at 150\% \cite{oasis}.} the user's vault is liquidated \cite{makerAuctions}.

Similar to research on the overall DeFi industry, formal research on Maker is relatively meager. Three relevant papers, all published in the past year, have focused on disparate topics concerning the project: potential exploits of Maker's governance voting process \cite{exploit1}, historical failures in Maker's pricing oracles \cite{mkrGovernance}, and a proposed model to evaluate default risk in Maker's loan portfolio \cite{mkrCredit}.

\subsection{Auction process}
In Maker's liquidation process, the user's collateral (``lot'') is put up for auction.\footnote{This type of auction, referred to as a ``collateral'' auction in the Maker process, is the focus of our analysis. The Maker process also employs ``surplus'' and ``debt'' auctions \cite{makerAuctions}. These auction types occur only rarely, and are out of scope for this paper.} The target proceeds (``tab'') include the value of the vault's debt, plus a liquidation penalty.\footnote{As of this writing, the liquidation penalty is set at 13\% \cite{oasis}.} All bids are submitted in DAI \cite{flipper}, and the bidders participating in the process are referred to as ``keepers'' \cite{keepers}.

The auction is completed in two parts. First, in the ``tend'' phase, the payment amount increases until the target proceeds are met. Second, in the ``dent'' phase, the reward received (the lot) decreases, until the auction reaches the maximum auction duration,\footnote{As of this writing, the maximum auction duration is set at 6 hours \cite{daiAuctions}.} or until no bidder is willing to bid lower than the current bid \cite{flipper}. Regardless of the auction phase, the progress of the auction can be measured through the auction price of the collateral, relative to the current market value of the collateral. 

The auction reward must be unlocked by submitting a ``deal'' transaction. If the auction reward is less than the collateral originally offered in the auction (as a result of decreasing bids in the dent phase), the difference is returned to the user owning the liquidated vault \cite{flipper}.

\begin{center}
    \begin{figure}[htb]
        \resizebox{\columnwidth}{!}{%
            \includegraphics{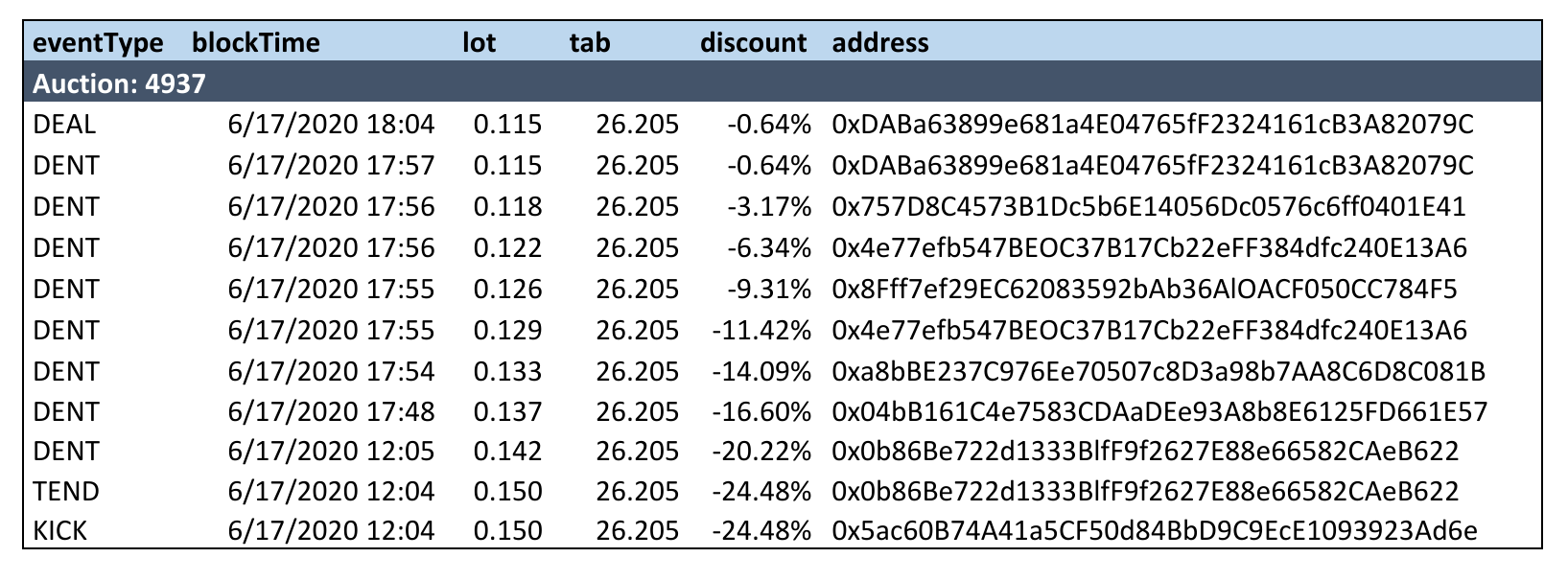}
        }
        \caption{Example auction results}
        \label{fig:auctionResults}
    \end{figure}
\end{center}

The results of a recent auction, completed in June 2020, are shown in Figure \ref{fig:auctionResults}.

The first liquidations to use this auction process were completed in November 2019.\footnote{Prior to November 2019, Maker used a fixed-discount collateral sale to complete liquidations \cite{scdLiquidations}; these sales are not considered in this paper, as they did not utilize an auction process.} Through July 31, 2020, approximately \$20.8M in collateral has been liquidated in this auction process \cite{liquidationVol}.

The most notable event in Maker's auction process occurred on March 12, 2020, when the price of ETH dropped in excess of 40\%. The rapid drop led to many vaults being liquidated after falling below the 150\% collateral ratio \cite{blackThursdayReport}. In total, almost 4,000 liquidations auctions were triggered on March 12, with a total value of approximately \$10.4M \cite{liquidationVol}.

Because of the high gas fees on the Ethereum network, many keepers were unable to submit bids on the auctions. Without a robust network of bidders, the handful of remaining bidders were able to submit zero-value bids. As a result, multiple vaults were liquidated at prices of zero, and the vaults' users did not receive any excess collateral \cite{blackThursdayReport}.

Further explanation of the auction process, in the context of formal auction theory, is given in Appendix \ref{section:characteristics}.

\section{Bidding strategy}
\label{sec:strategy}
\subsection{Conceptual example}
In order to define the bidding strategy for Maker auctions, we start with a conceptual example: bidding for a jar filled with $q$ quarters. The quarters have, in total, a single market value, which can be expressed in dollars as $0.25q$. The seller does not hide the amount of $q$; all bidders know the exact number of $q$, and therefore the total value of the jar of coins. If no other factors were present, it may be predicted that all bidders would bid exactly $0.25q$, because each bidder has exactly the same valuation.

However, we must consider two other components to the bidder valuations.
\subsubsection{Alternative-usage value} While quarters are valued at \$0.25 cents by all bidders, the coins could be worth more to certain individuals who could make additional profit with the coins. For example, the coins could be melted down and the raw materials used to produce another item that would be worth more than $0.25q$. The alternative-usage value of the coins, in excess of $0.25q$, is defined as $a$ for an individual bidder. In all future references, $a$ refers to the excess of the alternative-usage value over the market value, and not the alternative-usage value itself.
\subsubsection{Participation costs} 
\begin{itemize}[leftmargin=.4in]
    \item \textbf{Transaction costs}: In this scenario, bidders must pay a small fee to submit their bid, and another small fee to collect the jar if their bid wins. The total of these fees is defined as $b$.
    \item \textbf{Conversion costs}: Bidders pay for the jar of coins in cash, but they can carry only a limited amount of cash in their wallet. If several auctions took place at the same time, a bidder who won an auction would likely not have enough cash to participate in a second auction. In order to have enough cash on hand, a bidder would need to periodically visit a bank and exchange their coins for additional cash. The expense of visiting the bank (the cost of transportation, the value of lost time, etc.) are defined as $c$.
    \item \textbf{Cost of capital}: If bidders increased the amount of cash held in their wallets, they would be able to avoid visiting the bank as frequently as when they held smaller amounts of cash. However, holding cash in a wallet prevents bidders from earning interest on cash deposited at the bank. The interest foregone when withdrawing cash from the bank is the cost of capital, defined as $d$.
\end{itemize}

Taking those private value considerations into effect, the expected bid price for an individual bidder can be defined as
\begin{equation}
\label{eq:coins}
    0.25q + a - (b + c + d)
\end{equation}

In theory, the bidder with the highest value of $a - (b + c + d)$ would be able to bid the highest price for the jar of coins.

\subsection{Application to Maker auctions}
In Maker auctions, the ``jar of coins'' is the collateral being auctioned. The collateral has a clearly-defined market value, and the value is known by all bidders before the auction commences. 

The additional components of the bidder valuations are as follows:
\subsubsection{Alternative-usage value} Bidders can sell their collateral on the market, but they can also use the collateral to gain additional profits elsewhere on the blockchain. In theory, bidders could gain more profit by using the collateral than they could holding DAI or US dollars.
\subsubsection{Participation costs}
\begin{itemize}[leftmargin=.4in]
    \item \textbf{Transaction fees}: In order to submit a bid (tend or dent), collect winnings (deal), or execute any other transaction, bidders must pay a fee, known as ``gas fees'' on the Ethereum blockchain.
    \item \textbf{Conversion costs}: Bidders must convert the collateral they have won back into DAI, in order to continue participating in auctions. Because bidders bid in DAI and receive back collateral of a different currency, winning bidders will quickly run out of DAI. When converting collateral back to DAI, bidders must account for a loss in value when executing a trade on a decentralized exchange. This loss is referred to as ``slippage.''
    \item \textbf{Cost of capital}: Bidders must invest some amount of capital in holding DAI; the expected return on their investment is their cost of capital.
\end{itemize}

Applying Equation \eqref{eq:coins}, the expected auction winner would be the bidder with the highest $a - (b + c + d)$, where $a$ is the alternative-usage value of the collateral, and $b + c + d$ are the participation costs. 

Using this model, we would expect the following conditions to be true. If participation costs are higher than alternative-usage value, then the winning auction price would be below the market price. If the alternative-usage value and the participation costs are equal, the winning auction price would be at the market price. If the alternative-usage value is higher than participation costs, the winning auction price would not (as might be expected) rise above the market price. Instead, the winning auction price should be bounded by the market value, because the highest bidder would be able to obtain the collateral at market prices elsewhere, and would have no incentive to bid above the market price.

\subsection{Two-person bidding example}
We can apply our theory in a scenario with only two bidders, $\alpha$ and $\beta$, participating in a Maker auction. We assume there is no alternative-usage value of the collateral to the bidders. Therefore, the only factors relevant to the bidders are their participation costs. We also assume that $\alpha$'s and $\beta$'s participation costs are 2\% and 3.5\% of the collateral value, respectively.

We assume that $\alpha$ and $\beta$ will continue bidding the price higher\footnote{In the first auction phase (tend), bidders increase the payment amount. In the second auction phase (dent), bidders lower the reward proceeds, which effectively raises the price of the collateral. For simplicity, all bids will be referred to as raising the price, without reference to whether the payment is being increased or the reward is being decreased.} until the auction discount\footnote{If the bidding price is below the market price, the difference between the two prices, divided by the market price is referred to as a ``discount'' and is expressed as a negative percentage. If the bidding price is above the market price, the difference between the two prices, divided by the market price, is referred to as a ``markup`` and is expressed as a positive percentage.} is equal one of the bidder's participation costs. $\beta$ can only bid the price up to 96.5\% of the collateral value (a discount of -3.5\%), while $\alpha$ can bid up to 98\% of the collateral value (a discount of -2\%). Therefore, $\alpha$ should always win in an auction setting, because $\alpha$ can bid at a higher price than $\beta$, without suffering a loss.

\begin{center}
    \begin{figure}[htb]
        \resizebox{\columnwidth}{!}{%
            \includegraphics{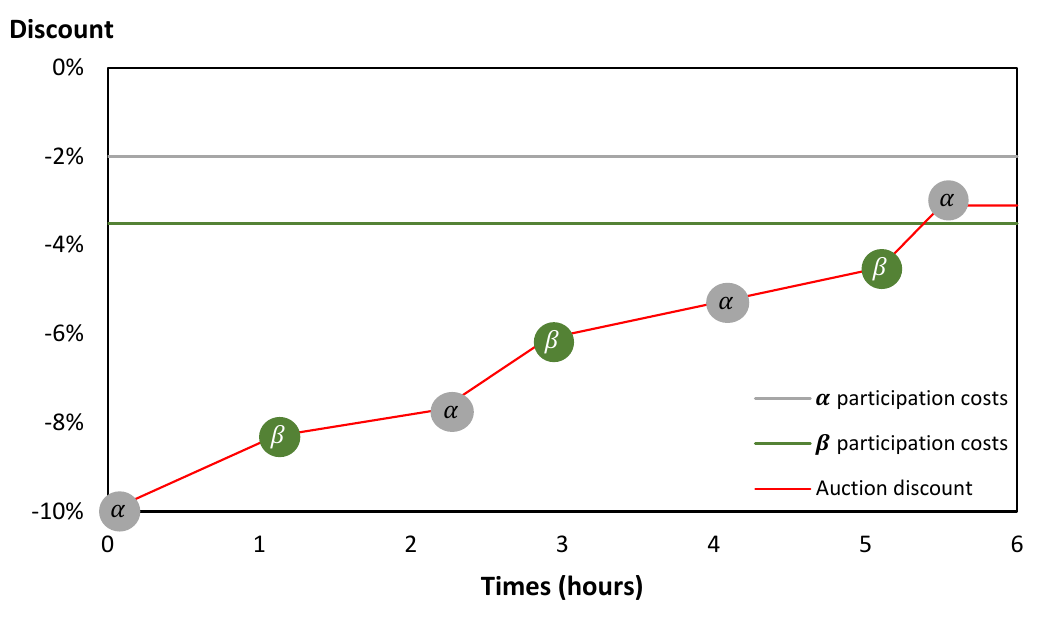}
        }
        \caption{Example auction with two bidders. Over the course of the 6-hour auction, the discount is reduced until it reaches the participation cost threshold of one of the bidders. At that point, the bidder with the lower threshold (the lower participation cost) will win the auction.}
        \label{fig:twoBidders}
    \end{figure}
\end{center}

\subsection{Proposed bidding strategy}
We first note that the value of alternative usage is a key component in determining the appropriate bid price. However, it is quite difficult to estimate the alternative-usage value for collateral. For purposes of this analysis, we focus primarily on participation costs, and only return to alternative-usage values at the end of this paper. 

With that caveat established, we can then propose an optimal bidding strategy. From the examples given above, we arrive at the following proposals:
\begin{enumerate}
    \item Participation costs determine bidding strategy.
    \item The bidder with the lowest participation costs will always win.
    \item Participation costs can be optimized to the lowest possible amount for each bid value.
\end{enumerate}

\begin{center}
    \begin{figure}[htb]
        \resizebox{\columnwidth}{!}{
            \includegraphics{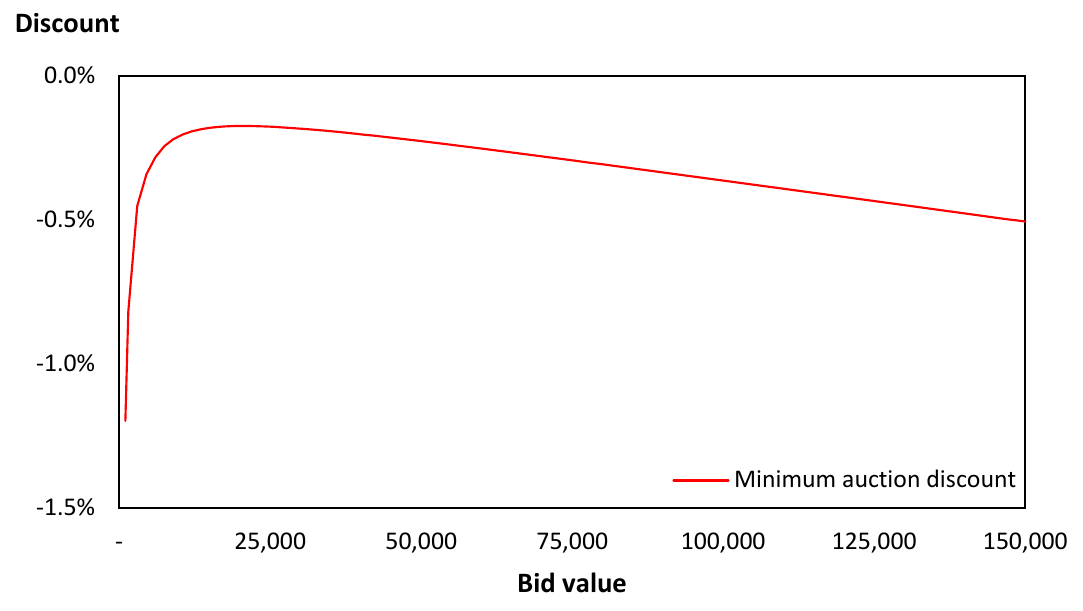}
        }
        \caption{Theoretical minimum auction discount}
        \label{fig:chart9}
    \end{figure}
\end{center}

By optimizing participation costs to their lowest possible total, a bidder could calculate the minimum participation costs required at every bid value. Dividing these costs into the market price yields a minimum auction discount, expressed as a percentage (see Figure~\ref{fig:chart9}). The minimum auction discount serves as a bidding threshold, because a bid with a discount smaller than the minimum auction discount (e.g. a higher price) would be unprofitable. Therefore, a bidder should never bid at a smaller discount than the minimum auction discount, or (in equivalent terms) should never bid at a price higher than the market price subtracted from participation costs.

\section{Participation costs}
\label{sec:costs}
Having established the importance of participation costs in determining bid valuations, we now define the specific costs incurred to participate in Maker auctions. 

In order to determine the costs incurred, we created a bot to run as a keeper on the Kovan test network. The Kovan network is a test version of the Ethereum network, and transactions executed on this network do not carry any monetary value. However, the functionality of the Kovan network closely mirrors the main Ethereum network. In addition, the Maker Protocol has set up identical smart contracts on both Kovan and the main Ethereum network, which allows for testing under near-real conditions on the Kovan network.

The keeper bot ran in May and June of 2020, and participated in a total of 42 auctions on the Kovan network. In addition to testing automated bidding, we also manually executed certain transactions, such as exchanging collateral rewards (WETH) for DAI, and vice-versa (see full transaction history at \cite{keeperBot}).

After analyzing the transactions required to run the bot, we identified three types of participation costs:
\begin{enumerate}
    \item \textbf{Transaction fees}: Gas fees to execute transactions on the Ethereum blockchain.
    \item \textbf{Conversion costs}: Slippage and trading fees for converting currencies on decentralized exchanges.
    \item \textbf{Cost of capital}: Implicit cost of holding capital in cryptocurrencies.
\end{enumerate}

The costs listed above could be calculated from onchain data, either directly (transaction fees and conversion costs) or indirectly (cost of capital). Although cost of capital is not directly incurred on the blockchain, its cost is derived from the value of the currency held onchain, and we therefore included the costs in our analysis. We did not include costs that were not directly related to on-chain transactions, such as the cost of hardware, electricity, maintenance, and other equipment costs. While these cost are incurred by keepers, we concluded that such costs were out-of-scope of our analysis, which focuses on the optimization of onchain costs. 

In the following sections, we define the specific components of participation costs.

\subsection{Transaction fees}
In Ethereum, charges for computing power are measured in ``gas'' $G$. Each transaction on Ethereum takes up a certain amount of gas. For example, a transfer between two non-smart contract addresses always takes up 21,000 $G$ \cite{gasUsage}. A transaction involving smart contracts would take up a greater amount of gas, with the exact value determined by the complexity of transaction.

Miners then charge a fee for the use of Ethereum's computational power. This fee is called the ``gas price'' $\mu$. The gas price is generally quoted in ``gwei'', which is worth \( 10^{-9} \) of a full unit of ETH (``ether''). The conversion factor between ether and gwei is represented throughout as \( g = 10^{-9} \).

The total transaction fee is calculated as the gas used multiplied by the gas price, or $G  \mu$. The resulting answer is in gwei; to convert to ethers, the largest unit of Ethereum, the answer can be calculated as $G \mu g$.

\subsubsection{Bid fees}
For purposes of this exercise, we assumed that bidders would not initiate (kick) an auction, and that bids would only be submitted in the dent phase (gas used $G_{dent}$).

If the auction is won, the rewards must be collected (gas used $G_{deal}$). $G_{deal}$ will only be incurred if the auction is won. Therefore, we also incorporate the probability $x$ of winning the auction. The total fee, expressed in terms of ETH, is 
\begin{equation}
    F_{bid} = (G_{dent} + G_{deal} x) \mu g 
\end{equation}

\subsubsection{Rebalance fees}
In the Maker system, DAI must be sent to a smart contract, called the Vat, before the DAI can be submitted for a bid. If a bid is won, the reward is received in WETH. Assuming auctions are won at a steady rate, the available balance of DAI in the Vat will be depleted over time, while the available balance of WETH in the Vat will increase over time.

In order to ensure a sufficient amount of collateral is available for bidding, the currency balance in the Vat must be periodically rebalanced, by converting WETH back to DAI. This operation cannot be performed inside the Vat. Therefore, the following three transactions are required, all of which require gas:
\begin{enumerate}
    \item Removing WETH from the Vat (\emph{exit})
    \item Converting WETH to DAI, using an exchange (\emph{trade})
    \item Adding DAI back to the Vat (\emph{join})
\end{enumerate}

The gas fees associated with these transactions are defined as $G_{exit}$, $G_{trade}$, and $G_{join}$, respectively. The three transactions described above do not need to be executed after every bid. Rather, their frequency depends on how quickly the Vat balance is depleted of the DAI needed to submit future bids, as well as the frequency of bids being successful.

It is assumed that a keeper will set a maximum amount of bidding capital in the Vat ($V_{max}$), and a minimum amount of capital ($V_{min}$). The difference between these two amounts is the ``rebalance margin'' $R$, or the amount of capital that is to be depleted before the Vat is rebalanced. 

Assuming the first transaction by a keeper sets the Vat capital at $V_{max}$, all subsequent bids will take up a certain percentage of $R$, before the capital needs to be rebalanced at $V_{min}$. Because the gas fees described above are only incurred when capital reaches $V_{min}$ (that is, when $R$ is fully depleted), each bid can be ascribed a proportional amount of gas fees, $y$, using the percentage the bid value $B$ is of $R$.

As an example, if $V_{max}$ is set at 10,000 DAI and $V_{min}$ is set at 7,500 DAI, then $R$ is 2,500. A $B$ of 1,000 DAI will represent 40\% of $R$. If gas fees, calculated and converted to DAI, are 0.2 DAI, then the gas fees ascribed to $B$ are 0.08 DAI (40\% x 0.2).

The proportional allocation does not, however, rise above 100\%. A $B$ of 5,000 DAI would be 200\% of an $R$ of 2,500. However, the rebalancing of the portfolio from WETH back to DAI would be performed in a single trade, not in multiple trades. Therefore, the proportional allocation $y$ would be calculated as 1 if $B \geq R$, and $\frac{B}{R}$ otherwise.

The allocated costs are further adjusted by the probability $x$ of winning the auction. Total rebalance fees are defined as 
\begin{equation}
     F_{rebal} = (G_{exit} + G_{trade} + G_{join}) \mu g x y
\end{equation}
\begin{equation}
\label{eq:4}
        y = 
        \begin{cases}
            1              &\quad \text{if } B\geq R\\
            \frac{B}{R}    &\quad \text{else}
        \end{cases}
\end{equation}

\subsubsection{Total gas fees}
The final step to calculating gas fees is to convert the amount in ethers to an amount in DAI. This conversion can be accomplished by multiplying by the WETH/DAI exchange rate. The exchange rate can be derived from the Uniswap reserves by calculating $\frac{T_0}{T_1}$ (see Section \ref{sec:conversion} for a detailed explanation of these values).
\begin{equation}
    \begin{split}
        F_{total} &= (F_{bid} + F_{rebal})  \frac{T_0}{T_1} \\
        & = (G_{dent} + G_{deal} x) \mu g \frac{T_0}{T_1} \\  
        & \quad + (G_{exit} + G_{trade} + G_{join}) \mu g x y \frac{T_0}{T_1} \\
    \end{split}
\end{equation}

\subsection{Conversion costs}
\label{sec:conversion}
As mentioned above, the account portfolio needs to be periodically rebalanced between WETH and DAI. This rebalance occurs by trading WETH for DAI on an exchange. To simplify our analysis, we assumed all trades were executed on Uniswap, the largest DeFi exchange by trading volume.\footnote{We note that Equations \eqref{eq:1} through \eqref{eq:3} are based primarily on the introductory work on Uniswap's mathematical characteristics by Angeris et al. \cite{defiMath1}.}

Uniswap uses a ``constant-product market-maker'' model, with ``liquidity pools'' set up as reserves for currency pairs. Given reserve $T_0$ for token 0, and reserve $T_1$ for token 1, the constant product $k$ will always equal \( T_0 T_1 \), in the absence of trading fees. 

Assuming that no trading fees are taken, the change \( \Delta T_0 \), caused by trading in an amount of $T_0$ to the liquidity pool, can be used to calculate the total change in the liquidity pool. Given the constant product nature of the liquidity pool, the new value of $k$ will be
\begin{equation} \label{eq:1}
    k = (T_1 - \Delta T_1)(T_0 + \Delta T_0)
\end{equation}

Rearranging the terms of Equation \eqref{eq:1}, the output amount of $T_1$ in a trade will be 
\begin{equation}
    \Delta T_1 = T_1 - \frac{k}{T_0 + \Delta T_0}
\end{equation}

When trading fees are introduced to the liquidity pool, $k$ now increases in proportion with trading fees $\gamma$, while still holding constant if fees are omitted from the equation.
\begin{equation}
\label{eq:2}
    k = (T_1 - \Delta T_1) (T_0 + (1-\gamma) \Delta T_0)
\end{equation}

Rearranging the terms of Equation \eqref{eq:2}, the output amount of $T_1$ in a trade will be 
\begin{equation}
    \Delta T_1 = T_1 - \frac{k}{T_0 + (1-\gamma) \Delta T_0}
\end{equation}

The implicit price in the liquidity pool before a trade, $P_0$, is given as \( P_0 = \frac{T_1}{T_0} \). The implicit price for the trade itself,  $P_1$, would be \( \frac{\Delta T_1}{\Delta T_0} \). Slippage $S$ is defined as the loss in value executed after the trade, and can be calculated as $\frac{P_0 - P_1}{P_0}$. In expanded form, $S$ is calculated as
\begin{equation}
    S = \frac{\frac{T_1}{T_0} - \frac{T_1 - \frac{T_0 T_1}{T_0 + (1-\gamma) \Delta T_0}}{\Delta T_0}}{\frac{T_1}{T_0}}
\end{equation}

After factoring the terms, $S$ may be expressed as
\begin{equation}
\label{eq:3}
    S = \frac{\gamma T_0 + (1-\gamma)\Delta T_0}{T_0 + (1-\gamma)\Delta T_0}
\end{equation}

In the context of Maker auctions, $\Delta T_0$ would normally be the amount of WETH exchanged for DAI. However, $\Delta T_0$ is dependent on the values of $B$ and $R$, both of which are expressed in terms of DAI. Therefore, we express $\Delta T_0$ as an amount of DAI, with reserve $T_0$ being the reserve for DAI. This conversion means that we are solving for slippage on the conversion of DAI to WETH, and not WETH to DAI, as would be the case in reality. However, the nature of constant-product markets is such that the slippage calculation results in the same answer, regardless of which currency is used as the input token.

If $B \geq R$, then an amount of WETH, equal in value to $B$, would need to be exchanged for DAI. Slippage would be calculated with $\Delta T_0$ equal to $B$ (in DAI), and the full slippage amount would be included in the cost calculation. If $B < R$, however, WETH would not need to be rebalanced until $R$ was fully depleted over multiple auctions. Therefore, when $B < R$, slippage would be calculated with $\Delta T_0$ equal to the $R$ (in DAI), and the resulting amount allocated to the cost calculation by multiplying by $\frac{B}{R}$. 

The conditional aspects of this calculation are represented by variables $y$ (allocation of costs up to 100\%) and $z$ (the value to use in the calculation, expressed in terms of DAI). $y$ is defined in Equation \eqref{eq:4}, and $z$ is defined as 
\begin{equation}
        z = 
        \begin{cases}
            B       &\quad \text{if } B\geq R\\
            R       &\quad \text{else}
        \end{cases}
\end{equation}

The amount of slippage is also adjusted by the probability $x$ of winning the auction (if the auction is not won, then no WETH will need to be exchanged). Finally, the slippage amount, which is in percentage form, is multiplied by $B$, so that slippage costs are expressed in terms of DAI. In expanded form, $S$ can be calculated as
\begin{equation}
\label{eq:5}
        S = \frac{\gamma T_0 + (1-\gamma)z}{T_0 + (1-\gamma)z} x y B
\end{equation}

\subsection{Cost of capital}
The amount of DAI capital held in the Vat is allowed to fluctuate between $V_{max}$ and $V_{min}$. However, the total portfolio value does not change, as any DAI used to pay for a bid is replaced by WETH of roughly the same value. Therefore, we assume that $V_{max}$ represents the average balance held throughout the year. $V_{max}$ is then subject to a capital charge $r$ (also known as the ``required rate of return''). The annual cost of capital is defined as 
\begin{equation}
    K_{annual} = r V_{max}
\end{equation}

$K_{annual}$ can then be allocated to an individual bid, assuming a certain number of bids in a year, $B_{year}$. Allocating equally to individual bids inherently assumes a constant rate of depletion throughout the year, including auctions won or lost. Because of this assumption, we do not need to explicitly include an adjustment for probability $x$ of winning an auction. 

In expanded form, the cost of capital ascribed to an individual bid would be
\begin{equation}
    \begin{split}
        K_{bid} = \frac{r V_{max}}{B_{year}}
    \end{split}
\end{equation}

\subsection{Total participation costs}
Total participation costs are defined as
\begin{equation}
\label{eq:totalCost}
    \begin{split}
        C &= F_{total} + S + K_{bid}
    \end{split}
\end{equation}

We measure $C$ relative to $B$, as $\frac{C}{B}$. Over smaller values of $B$, $C$ deceases as a percentage of $B$, because of the fixed and semi-fixed nature of $K_{bid}$ and $F_{total}$, respectively. At higher values of B, however, $C$ increases as a percentage of $B$, because of the increasing costs of $S$.

\section{Optimization}
\label{sec:optimize}
\subsection{Optimization and constraints}

In Equation \eqref{eq:totalCost}, the two terms controllable by a bidder are $V_{max}$ (maximum portfolio value) and $R$ (rebalance margin). We undertook to solve for the lowest possible participation costs, using auction-specific data, by adjusting the values for $V_{max}$ and $R$. 

We selected an appropriate time period to analyze (see further details in Section \ref{sec:params}), during which 155 auctions were completed in the Maker system. We selected the final winning bid for each auction and then ran our optimization 155 times, with parameters derived from the winning bid for each auction. We then compared the theoretical minimum auction discount, as recommended by our optimization, to the actual discounts in our historical data (see Section \ref{sec:results}).

The optimization was bounded by several constraints, in order to mirror real conditions:

\begin{itemize}
    \item The minimum value of the portfolio ($V_{max}-R$) must be enough to cover the assumed bid value $B$. This constraint also ensures that $V_{max}$ is greater than $R$, which is necessary because $R$ is subtracted from $V_{max}$, and the resulting value cannot be negative.
    \item The maximum portfolio value and rebalance margin cannot be negative (infeasible), and the maximum portfolio value cannot be zero (this would signify non-participation).
\end{itemize}

The optimization is shown below in its full form. 
\begin{equation}
    \begin{split}
        \min_{V_{max}, R} \quad &(G_{dent} + G_{deal} x) \mu g \frac{T_0}{T_1} \\  
        &+(G_{exit} + G_{trade} + G_{join}) \mu g x y \frac{T_0}{T_1} \\
        &+\frac{\gamma T_0 + (1-\gamma)z}{T_0 + (1-\gamma)z} x y B\\
        &+\frac{r V_{max}}{B_{year}} \\
         &y = 
        \begin{cases}
            1              \quad \text{if } B\geq R\\
            \frac{B}{R}    \quad \text{else}
        \end{cases}
        \\
        &z = 
        \begin{cases}
            B       \quad \text{if } B\geq R\\
            R       \quad \text{else}
        \end{cases} \\
        \text{s.t.} \quad &V_{max} - R \geq B \\
        &V_{max} > 0 \\
        & R \geq 0
    \end{split}
\end{equation}

The objective function has two branches, depending on the relation of $B$ and $R$. Considered independently, the functions resulting from the two branches are both convex. As we desired to know the values of $R$ and $V_{max}$ that minimize the cost for a given value of $B$, we found the optimal of the two branches separately and then chose the $R$ and $V_{max}$ corresponding to the minimum cost of the two as the answer. The implementation was completed in MATLAB.

\subsection{Parameters and data collection}
\label{sec:params}
The full list of parameters used in our optimization is included below. Further explanation of these parameters is given in the following sections.

\begin{table}[htb]
    \caption{Optimization parameters}
    \label{table1}
    \resizebox{\columnwidth}{!}{%
        \begin{tabular}{|l | l | l| l|} 
            \hline
            \multicolumn{4}{|c|}{\textbf{\textit{Auction-specific}}} \\ 
            \hline
            \it Param & \it Value & \it Definition & \it Source  \\
            \hline
            $B$ & Variable & Value of winning auction bid & (1) \\ 
            $\mu$ & Variable & Gas price at time of bid & (1) \\ 
            $T_0$ & Variable & DAI reserve for ETH-DAI Uniswap pair & (2) \\ 
            $T_1$ & Variable & ETH reserve for ETH-DAI Uniswap pair & (2) \\
            \hline
            \multicolumn{4}{c}{ } \\
            \hline
            \multicolumn{4}{|c|}{\textbf{\textit{Transaction fees}}} \\
            \hline
            \it Param & \it Value & \it Definition & \it Source  \\
            \hline
            $G_{dent}$ & 116,914 & Gas used to submit dent transaction & (1)\\ 
            $G_{deal}$ & 44,154 & Gas used to submit deal transaction & (1) \\
            $G_{exit}$  & 80,145 & Gas used to submit exit transaction & \cite{exitJoinData} \\ 
            $G_{trade}$ & 125,700 & Gas used to submit trade transaction & \cite{tradeData} \\ 
            $G_{join}$ & 80,380 & Gas used to submit join transaction & \cite{exitJoinData} \\ 
            $g$ & $10^{-9}$ & Conversion from gwei to ethers & (3) \\ 
            $\gamma$ & 0.003 & Uniswap trading fee for analysis period & \cite{uniFees} \\ 
            \hline
            \multicolumn{4}{c}{ } \\
            \hline
            \multicolumn{4}{|c|}{\textbf{\textit{Cost of capital}}} \\ 
            \hline
            \it Param & \it Value & \it Definition & \it Source  \\
            \hline
            $B_{year}$ & 365 & Number of bids per year & (1) \\ 
            $r$ & 40\% & Cost of capital for cryptocurrency & (4) \\
            \hline
            \multicolumn{4}{c}{ } \\
            \hline
            \multicolumn{4}{|c|}{\textbf{\textit{Other}}} \\
            \hline
            \it Param & \it Value & \it Definition & \it Source  \\
            \hline
            $x$ & 15\% & Win probability for individual bid & (1) \\
            \hline
        \end{tabular}
    }
    \vspace{0.1cm}
    
    (1) Maker auction data \\
    (2) Uniswap trading data \\
    (3) Ethereum specifications \\
    (4) Previous valuations \\

    \caption{Optimization variables}
    \label{table2}
    \resizebox{\columnwidth}{!}{%
        \begin{tabular}{|l | l | l|}
            \hline
            \multicolumn{3}{|c|}{\textbf{\textit{Objectives}}} \\
            \hline
            \it Param & \it Value & \it Definition  \\
            \hline
            $V_{max}$ & Variable & Maximum value of portfolio \\ 
            $R$ & Variable & Amount of depletion in $V_{max}$ before rebalancing \\
            \hline
        \end{tabular}
    }
\end{table}

\subsubsection{Auction-specific  data}
We began by downloading all Maker auction events (kick, tend, dent, and deal) from the relevant Maker smart contract on the Ethereum blockchain \cite{makerHistory}. All data was downloaded through an Infura node, which we queried using NodeJS running on an Ubuntu 18.04 virtual machine.\footnote{All code used in this paper may found at \url{https://github.com/michael-darlin/optimal-bidding-strategy}.}

We downloaded the auction events from the beginning of the Maker liquidation process (November 13, 2019) through the date that the Maker auction process was upgraded to use new smart contracts (July 28, 2020). Our analysis focused specifically on the period of March 23, 2020 through July 28, 2020. The beginning date of March 23 was chosen in order to exclude outliers, such as the zero-value bids of March 12, 2020, and a handful of bids that were submitted at prices many times higher than the prevailing market price (likely in error). The end date of July 28, 2020 was chosen so that the auction process would be consistent across all auctions analyzed.

We also downloaded information on the Uniswap reserves for the ETH-DAI trading pair from Uniswap's GraphQL node \cite{uniswapV1} (for version 1 of the Uniswap protocol), and from the Uniswap smart contract for the ETH-DAI trading pair on the Ethereum blockchain \cite{uniswapV2} (for version 2 of the protocol).

From this data set, we were able to derive the auction-specific parameters, which changed from auction to auction based on the auction settings or the conditions of the Ethereum network.

We note that Uniswap upgraded their protocol from version 1 to version 2 on May 19, 2020. Since the upgrade, both version 1 and version 2 have had active ETH-DAI trading pairs, albeit with the majority of the volume shifting to version 2 over time. In our optimization, we set $T_0$ and $T_1$ equal to the reserves of whichever pair had the larger reserves. In practice, this condition meant that the trading pair for version 1 was used through June 25, 2020, and the trading pair for version 2 was used thereafter.

\subsubsection{Transaction fees} Because of the complex calculations required to estimate the gas usage for transactions involving smart contracts, we used historical data to estimate the typical gas used for each Maker transaction type (dent, deal, exit, and join), as well as for Uniswap transactions (trade). For each event type, we selected the most recent 50 transactions (except for exit and join, which were downloaded as one set of 50 transactions), either from our database of auction events or from the transaction history publicly available on Etherscan. This data was downloaded between June 20 and June 22, 2020.

We found that gas usage was higher for multi-step transactions, in which multiple events were executed in a single transaction. We assumed that bidders would execute transactions step-by-step; therefore, we considered only transactions with a single event being executed. Within the group of single-event transactions, we chose the gas usage with the most frequent occurrence in the data. If multiple values had the highest occurrence, we chose the highest amount of gas usage.

We note that the average cost for trading tokens varied depending on the number of tokens involved; some trades could involve trading from Currency A to B (two tokens), or Currency A to Currency B to Currency C (three tokens), and so on. In our analysis, we collected data from trades using up to four tokens, and then used a regression model to estimate the gas fees used for trading. When $t$ tokens were involved in a trade, the gas used for trading ($G_{trade}$) was estimated at $(47,912 t) + 29,876$. For simplicity, two tokens were assumed to be involved in each trade, which led to a value of 125,700 for $G_{trade}$.

\subsubsection{Cost of capital}
The allocation of the cost of capital depends on the number of bids made in the year ($B_{year}$). In the selected data set, 155 auctions were completed over the period of 128 days (March 23 through July 28), a rate of approximately 1.2 auctions per day. We rounded this number to 1 bid per day, resulting in a value of 365 for $B_{year}$. 

The allocated cost also depends on the cost of capital percentage $r$, also known as the discount rate. Because of the uncertain nature of their future utility, cryptocurrencies are generally valued using very high discount rates. While no single number can be defined as the appropriate discount rate, we observed that most valuations used discount rates between 30\% at the low end \cite{valuation1} and 50\% at the high end \cite{valuation2}. From this range, we chose the mid-point value of 40\%, a rate which has itself been used in several prior valuations \cite{valuation3, valuation4}. A discount rate of 40\% also falls within the range of returns expected for startup companies that are growing but still unprofitable \cite{valuation5}, a description applicable to many cryptocurrency projects.

We note that the valuations referenced above assume cryptocurrencies have highly volatile prices when compared to the US dollar. To our knowledge, formal research has not examined the valuation of stablecoins, such as the DAI currency that we assume is being held in a bidder's portfolio. Stablecoins are designed to be pegged to the US dollar, and in theory could use a discount rate that approaches the risk-free rate. In practice, however, stablecoins have many riskful characteristics that would increase their risk premium above the risk-free rate. Defining a stablecoin-specific discount rate is outside the scope of this paper, and for purposes of this analysis, we align our discount rate with those used by prior cryptocurrency valuations.

\subsubsection{Other} We calculated win probability $x$ by aggregating all bids associated to the auctions included in our analysis (two auctions were initiated on March 22, but concluded on March 23, which resulted in our history extending back to March 22). We then calculated the total number of bids (1,011). Dividing the total number of winning bids by the total number of bids resulted in a value of approximately 15\% for $x$.

\subsection{Optimization results}
\label{sec:results}
After optimizing the participation costs in 155 auctions, we compared our optimal bidding price to the actual auction-winning bid prices.

\begin{table}[htb]
    \caption{Auction sample characteristics}
    \centering
    \ra{1.3}
        \begin{tabular}{@{}lcc@{}}
            \toprule
            \bf Bid value & \bf Count & \bf \% \\ 
            \midrule
            \$1 - \$1,000 & 114 & 74 \\
            \$1,001 - \$10,000 & 18 & 12 \\ 
            > \$10,000 & 23 & 15 \\
            \hline
            All & 155 & 100 \\
            \bottomrule
        \end{tabular}
\end{table}

\begin{table}[htb]
    \caption{Optimization results}
    \centering
    \ra{1.3}
    \resizebox{\columnwidth}{!}{%
        \begin{tabular}{@{}lcccccc@{}}
            \toprule
            \bf Bid value & \phantom{n} & \multicolumn{2}{c}{\bf Actual > Optimal} & \phantom{n} & \multicolumn{2}{c}{\bf Actual < Optimal} \\
            \cmidrule{3-4} \cmidrule{6-7}
            && Count & \% && Count & \% \\ 
            \midrule
            \$1 - \$1,000 && 95 & 83 && 19 & 17 \\
            \$1,001 - \$10,000 && 10 & 56 && 8 & 44 \\ 
            > \$10,000 && 12 & 52 && 11 & 48 \\
             \cmidrule{1-7}
            All && 117 & 75 && 38 & 25 \\
            \bottomrule
        \end{tabular}
    }
\end{table}

We first observed that the actual bidding price was higher than the optimal bidding price in 75\% of the auctions. While this condition was true for over 80\% of auctions with a winning bid value equal to or less than \$1,000, it was true for only slightly more than half of auctions with a bid value greater than \$1,000.

\begin{center}
    \begin{figure}[htb]
        \resizebox{\columnwidth}{!}{
            \includegraphics{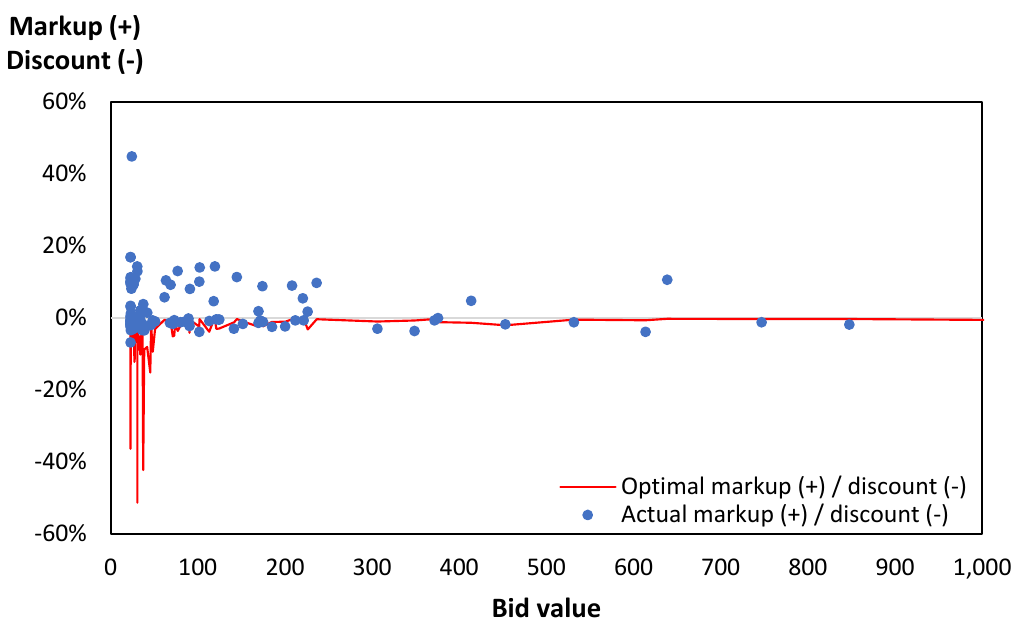}
        }
        \caption{Comparison of optimal markup or discount vs. actual markup or discount, for all bids up to \$1,000}
        \vspace{0.4cm}
        \resizebox{\columnwidth}{!}{
            \includegraphics{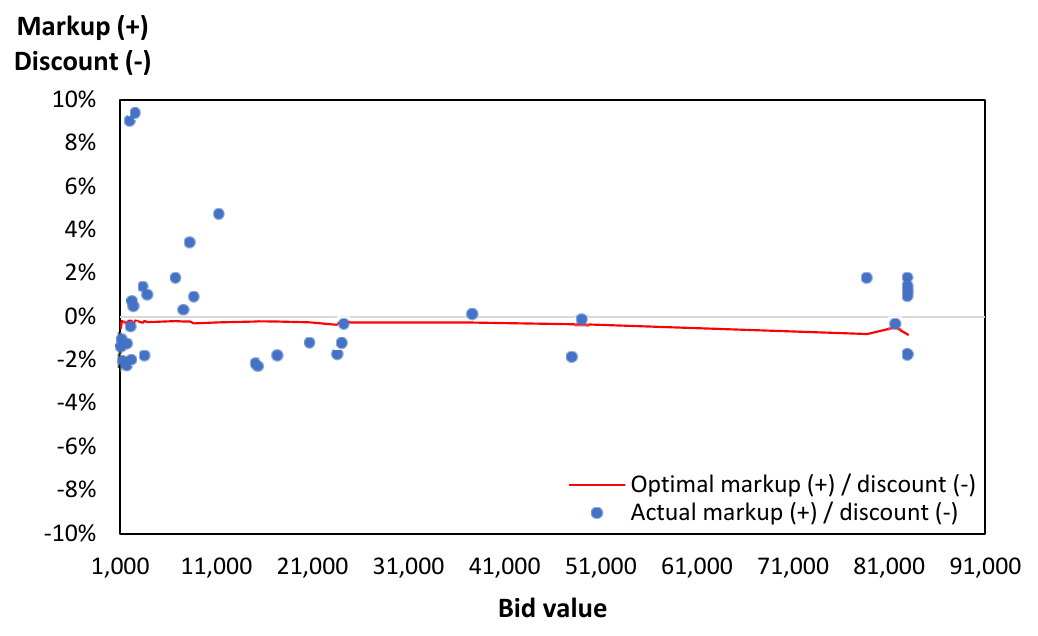}
        }
        \caption{Comparison of optimal markup or discount vs. actual markup or discount, for all bids over \$1,000}
    \end{figure}
\end{center}

It may be theorized that individual bidders did not include cost of capital in their calculations, because the cost is implicit only. If participation costs only included explicit costs (transaction fees and conversion costs), then the resulting minimum auction discount may be closer to the actual auction discount. Therefore, we modified the calculation of total participation costs to exclude cost of capital, with results shown in Table~\ref{tab:2ndRound}.
\begin{table}[htb]
    \caption{Optimization results, with cost of capital excluded}
    \centering
    \ra{1.3}
    \resizebox{\columnwidth}{!}{%
        \begin{tabular}{@{}lcccccc@{}}
            \toprule
            \bf Bid value & \phantom{n} & \multicolumn{2}{c}{\bf Actual > Optimal} & \phantom{n} & \multicolumn{2}{c}{\bf Actual < Optimal} \\
            \cmidrule{3-4} \cmidrule{6-7}
            && Count & \% && Count & \% \\ 
            \midrule
            \$1 - \$1,000 && 93 & 82 && 21 & 18 \\
            \$1,001 - \$10,000 && 10 & 56 && 8 & 44 \\ 
            > \$10,000 && 12 & 52 && 11 & 48 \\
             \cmidrule{1-7}
            All && 115 & 74 && 40 & 26 \\
            \bottomrule
        \end{tabular}
    }
    \label{tab:2ndRound}
\end{table}

This modification changed the results of only two auctions, leaving the majority of bids still at a price higher than optimal, with most above-optimal bids occurring in auctions with bid values of \$1 to \$1,000.

\subsection{Discussion}
\label{sec:discussion}
From the results of the analysis above, it is evident that the majority of auction-winning bids were at prices that would not allow bidders to recoup their participation costs. Our discussion of this seemingly unprofitable behavior begins by considering two potential reasons, which are ultimately rejected as feasible explanations. We then describe three reasons that may serve as probable explanations for this behavior.
\newline

\noindent \textbf{Reasons not accepted}
\subsubsection{Infeasibility of optimal portfolio to individual bidders} We acknowledge that it would be infeasible for bidders to adjust their portfolio size to be optimal at every value of $B$. The bid value of each new auction cannot be known in advance; in addition, multiple auctions with different values can be triggered at the same time. In hindsight, we were able to calculate what would have been the optimal portfolio size; in practice, however, it is impossible to adjust the value of $V_{max}$ and $R$ to arrive at the optimal cost for every new auction. Therefore, bidders would need to adjust their portfolio values to be optimal for just one value of $B$.

However, even if an individual bidder is unable to arrive at the optimal price for every auction, the totality of bidders participating in an auction should reach a near-optimal price for each auction. For example, Bidder A may be optimized for a $B$ of \$10,000, Bidder B for a $B$ of \$5,000, Bidder C for a $B$ of \$1,000, and so on. With multiple bidders optimized for a range of $B$ values, each auction should have a winning price that approaches the optimal price for that auction. Therefore, we do not believe that the infeasibility of an optimal portfolio for an individual bidder explains the gap in optimal versus actual prices.

\subsubsection{Use of auctions as a trading mechanism} It may be conjectured that bidders are not interested in making profits, but rather in exchanging DAI for WETH cheaply, which can be accomplished through the auction process. However, the Maker auction process requires multiple transactions and a wait of up to several hours before auction collateral can be collected. In contrast, decentralized exchanges allow DAI and WETH to be traded nearly instantaneously in a single transaction. As a result, we believe it is unlikely that individuals would use the Maker auction process, in its current form, as a trading mechanism.
\newline


\setcounter{subsubsection}{0}

\noindent \textbf{Proposed reasons}
\subsubsection{Indifference to cost of capital} The cost of capital is an implicit cost that does not appear on a transaction record or a wallet balance. Therefore, some bidders may disregard this cost when drawing up their bidding strategy. This condition may be particularly true for bidders who hold cryptocurrency based on personal preference (such as to avoid using money in the traditional financial system), rather than as a financial investment. For these bidders, the theoretical required return for cryptocurrencies may be of little consequence in their day-to-day decision-making.

\subsubsection{Inexperienced actors} Although some bidders may be indifferent to cost of capital, no bidder should be indifferent to explicit onchain costs, such as transaction fees and conversion costs.  However, as shown in Table \ref{tab:2ndRound}, three-fourths of bids submitted did not cover transactions fees and conversion costs. These results indicate certain bidders may not be aware of the full costs that are required to participate in Maker auctions. Although running an automated keeper bot requires a high degree of technical sophistication, we cannot dismiss the possibility that some bidders may devise their bidding strategy without a comprehensive accounting of the requisite costs.

\subsubsection{Altruistic actors} Indifference to cost of capital or lack of experience may explain why certain bids do not cover all participation costs; these theories do not explain why bids which are submitted at above-market prices. Any bidders that submit bids at prices higher than the market price are guaranteed to experience a loss on their portfolio value, even before subtracting participation costs.
    
Although these bids may not be rational from a financial standpoint, they indicate the presence of other non-financial motivations. Certain bidders may be motivated to strengthen the Maker ecosystem as a whole and prevent disruptive events, such as the zero-value bids of March 12, 2020. These bidders can therefore lose money on a single bid, but still profit through the smooth running of the system overall. Returning to our conceptual ``jar of coins'' model, certain bidders may have an alternative value $a$ for the collateral, which is the guarantee that a smooth auction process secures the stability of the system overall. These bidders can bid above the market price, because their total profit of $a$ (alternative-usage) $-(b+c+d)$ (participation costs) is positive. The value of $a$ is a private-value component for each bidder, in what is otherwise a common-value auction (as discussed in Appendix \ref{section:characteristics}).
    
The identities of these altruistic bidders are unknown in the anonymous setting of the Ethereum blockchain. However, these bidders could include any individual or organization with an incentive to ensure the Maker system runs smoothly.


\section{Conclusion}
\label{sec:conclusion}
In this paper, we have proposed an optimal bidding strategy for Maker auctions, based on minimizing the costs of participation. When comparing the proposed optimal bidding price to historical data, we find that the majority of auctions were won at prices higher than the optimal bidding price. 

We can suggest three avenues through which this research can be further extended. First, the optimal bidding price may be modified by including additional factors that influence bidding behavior. Potential factors to consider include the time at which the bid is placed in the auction lifecycle, the number of bidders participating, and external conditions on the Ethereum blockchain.

Second, our paper focused on prices above the optimal price, and did not explore in detail why a quarter of auctions finished at prices below the optimal price. Further research may uncover why certain auctions finish at prices that allow for bidder profits, while many others do not.

Finally, the theoretical model will require modifications under the newly proposed Maker auction system. This system has not been formally specified, but a preliminary proposal outlines the use of a Dutch auction system, in which bid prices start high and gradually decrease over time \cite{systemRedesign}. This change would allow bids to be won in a single transaction, which opens the possibility of using ``flash loans'' to bid on auctions without pre-existing capital. The new auction system will undoubtedly change the optimal strategy for bidders, and will provide a fresh area of research once the new process has been fully implemented.

\bibliography{references} 

\begin{thebibliography}{10}
\providecommand{\url}[1]{#1}
\csname url@samestyle\endcsname
\providecommand{\newblock}{\relax}
\providecommand{\bibinfo}[2]{#2}
\providecommand{\BIBentrySTDinterwordspacing}{\spaceskip=0pt\relax}
\providecommand{\BIBentryALTinterwordstretchfactor}{4}
\providecommand{\BIBentryALTinterwordspacing}{\spaceskip=\fontdimen2\font plus
\BIBentryALTinterwordstretchfactor\fontdimen3\font minus
  \fontdimen4\font\relax}
\providecommand{\BIBforeignlanguage}[2]{{%
\expandafter\ifx\csname l@#1\endcsname\relax
\typeout{** WARNING: IEEEtran.bst: No hyphenation pattern has been}%
\typeout{** loaded for the language `#1'. Using the pattern for}%
\typeout{** the default language instead.}%
\else
\language=\csname l@#1\endcsname
\fi
#2}}
\providecommand{\BIBdecl}{\relax}
\BIBdecl

\bibitem{ethWhitePaper}
V.~Buterin, ``{Ethereum Whitepaper},''
  \url{https://ethereum.org/en/whitepaper/}, accessed on Aug 12, 2020.

\bibitem{wsjDefi}
P.~Vigna, ``{Bitcoin Is Riding High Again as Investors Embrace Risk},''
  \emph{Wall Street Journal}, August 2020,
  \url{https://www.wsj.com/articles/bitcoin-is-riding-high-again-as-investors-embrace-risk-11596376800}.

\bibitem{wsjStablecoins}
T.~Geron, ``{Why Stablecoins Stand Out in the Cryptocurrency World},''
  \emph{Wall Street Journal}, June 2019,
  \url{https://www.wsj.com/articles/why-stablecoins-stand-out-in-the-cryptocurrency-world-11560218460/}.

\bibitem{defiPulse}
{DeFiPulse}, ``{DeFi - The Decentralized Finance Leaderboard at DeFi Pulse},''
  \url{https://defipulse.com/}, accessed on Aug 3, 2020.

\bibitem{etoro}
{eToro and The TIE}, ``{The State of Digital Assets Q2 2020},''
  \url{https://thetie.io/etoro-q2-2020-state-of-digital-assets}, accessed on
  Aug 3, 2020.

\bibitem{consensys.Q2Defi}
{Consensys}, ``{The Q2 2020 DeFi Report},''
  \url{https://consensys.net/insights/q2-2020-defi-report/}, accessed on Aug 3,
  2020.

\bibitem{defiOverview1}
D.~A. Zetzsche, D.~W. Arner, and R.~P. Buckley, ``{Decentralized Finance
  (DeFi)},'' \emph{European Banking Institute Working Paper Series 59/2020},
  March 2020.

\bibitem{defiOverview2}
F.~Schär, ``{Decentralized Finance: On Blockchain- and Smart Contract-based
  Financial Markets},'' \url{http://dx.doi.org/10.2139/ssrn.3571335}, March
  2020.

\bibitem{exploit1}
L.~Gudgeon, D.~Perez, D.~Harz, A.~Gervais, and B.~Livshits, ``{The
  Decentralized Financial Crisis: Attacking DeFi},'' in \emph{{2020 Crypto
  Valley Conference}}, June 2020.

\bibitem{exploit2}
K.~Qin, L.~Zhou, B.~Livshits, and A.~Gervais, ``{Attacking the DeFi Ecosystem
  with Flash Loans for Fun and Profit},''
  \url{https://arxiv.org/abs/2003.03810}, March 2020.

\bibitem{defiMath1}
G.~Angeris, H.-T. Kao, R.~Chiang, C.~Noyes, and T.~Chitra, ``{An analysis of
  Uniswap markets},'' in \emph{{2020 Cryptoeconomics System}}, March 2020.

\bibitem{defiMath2}
G.~Angeris and T.~Chitra, ``{Improved Price Oracles: Constant Function Market
  Makers},'' \url{https://arxiv.org/abs/2003.10001}, June 2020.

\bibitem{makerWhitePaper}
{Maker Foundation}, ``{Maker - Whitepaper},''
  \url{https://makerdao.com/en/whitepaper/}, accessed on Aug 28, 2020.

\bibitem{oasis}
------, ``{Oasis},'' \url{https://oasis.app/borrow/}, accessed on Aug 12, 2020.

\bibitem{makerAuctions}
------, ``{The Auctions of the Maker Protocol},''
  \url{https://docs.makerdao.com/auctions/the-auctions-of-the-maker-protocol/},
  accessed on Aug 22, 2020.

\bibitem{mkrGovernance}
W.~Gu, A.~Raghuvanshi, and D.~Boneh, ``{Empirical Measurements on Pricing
  Oracles and Decentralized Governance for Stablecoins},''
  \url{http://dx.doi.org/10.2139/ssrn.3611231}, June 2020.

\bibitem{mkrCredit}
A.~Evans, ``{A Ratings-Based Model for Credit Events in MakerDAO},''
  \url{https://static1.squarespace.com/static/5a479ee3b7411c6102f75729/t/5d37587d026881000198ef51/1563908221879/Maker-Ratings.pdf},
  July 2019.

\bibitem{flipper}
{Maker Foundation}, ``{Flipper - Detailed Documentation},''
  \url{https://docs.makerdao.com/smart-contract-modules/collateral-module/flipper-detailed-documentation},
  accessed on Aug 12, 2020.

\bibitem{keepers}
------, ``{Auction Keepers},''
  \url{https://docs.makerdao.com/keepers/auction-keepers}, accessed on Aug 14,
  2020.

\bibitem{daiAuctions}
{Dai Auctions}, ``{MCD Collateral Auctions},'' \url{https://daiauctions.com/},
  accessed on Aug 12, 2020.

\bibitem{scdLiquidations}
{Maker Foundation}, ``{Liquidation (SCD)},''
  \url{https://community-development.makerdao.com/makerdao-scd-faqs/scd-faqs/liquidation},
  accessed on Aug 12, 2020.

\bibitem{liquidationVol}
{DeBank}, ``{Liquidation data},''
  \url{https://cached-api.dappub.com/defi-insight/debt/all/liquidates},
  accessed on Aug 12, 2020.

\bibitem{blackThursdayReport}
{PeckShield}, ``{Black Thursday for MakerDAO: \$8.32 million was liquidated for
  0 DAI},''
  \url{https://medium.com/@whiterabbit_hq/black-thursday-for-makerdao-8-32-million-was-liquidated-for-0-dai-36b83cac56b6},
  March 2020.

\bibitem{keeperBot}
Etherscan, ``{Address Information},''
  \url{https://kovan.etherscan.io/address/0xd536ea64b9865059fc5e2d8bfd9aa9bf677722f3},
  accessed in August 2020.

\bibitem{gasUsage}
{Ethereum Wiki}, ``{Design Rationale},''
  \url{https://eth.wiki/en/fundamentals/design-rationale}, accessed on Aug 22,
  2020.

\bibitem{exitJoinData}
Etherscan, ``{Address Information},''
  \url{https://etherscan.io/address/0x9759A6Ac90977b93B58547b4A71c78317f391A28},
  accessed in June 2020.

\bibitem{tradeData}
------, ``{Address Information and Transactions},''
  \url{https://etherscan.io/address/0x7a250d5630b4cf539739df2c5dacb4c659f2488d},
  accessed in June 2020.

\bibitem{uniFees}
Uniswap, ``{Fees},'' \url{https://uniswap.org/docs/v2/advanced-topics/fees},
  accessed on Aug 20, 2020.

\bibitem{makerHistory}
Etherscan, ``{Address Information},''
  \url{https://etherscan.io/address/0xd8a04F5412223F513DC55F839574430f5EC15531},
  accessed in July 2020.

\bibitem{uniswapV1}
{The Graph}, ``{Uniswap Subgraph},''
  \url{https://thegraph.com/explorer/subgraph/graphprotocol/uniswap}, accessed
  in August 202.

\bibitem{uniswapV2}
Etherscan, ``{Address Information},''
  \url{https://etherscan.io/address/0xA478c2975Ab1Ea89e8196811F51A7B7Ade33eB11},
  accessed in August 2020.

\bibitem{valuation1}
R.~Kyburz, ``{Cryptoasset Research - Bitcoin},''
  \url{https://blocknovum.com/wp-content/uploads/2018/08/BlockNovum_Investment-Research_Bitcoin-August2018.pdf},
  August 2018.

\bibitem{valuation2}
``{Fundamental Valuation of Cryptoassets},''
  \url{https://drwvc.com/documents/2018-08-DRW-VC-Fundamental-Valuation-of-Cryptoassets.pdf},
  August 2018.

\bibitem{valuation3}
S.~Dowlat, ``{Cryptoasset Market Coverage Initiation: Valuation},''
  \url{https://research.bloomberg.com/pub/res/d37g1Q1hEhBkiRCu_ruMdMsbc0A},
  August 2018.

\bibitem{valuation4}
C.~Burniske, ``{Cryptoasset Valuations},''
  \url{https://medium.com/@cburniske/cryptoasset-valuations-ac83479ffca7},
  September 2017.

\bibitem{valuation5}
A.~Damodaran, ``{Valuing Young, Start-up and Growth Companies: Estimation
  Issues and Valuation Challenges},''
  \url{http://dx.doi.org/10.2139/ssrn.1418687}, May 2009.

\bibitem{systemRedesign}
{Maker Foundation}, ``{A Liquidation System Redesign: A Pre-MIP Discussion},''
  \url{https://forum.makerdao.com/t/a-liquidation-system-redesign-a-pre-mip-discussion/2790},
  June 2020.

\bibitem{auctionTheory.Klemperer}
P.~Klemperer, \emph{{Auctions: Theory and Practice}}.\hskip 1em plus 0.5em
  minus 0.4em\relax Princeton, NJ: Princeton University Press, 2004.

\bibitem{auctionTheory.Milgrom}
P.~R. Milgrom, \emph{{Putting Auction Theory To Work}}.\hskip 1em plus 0.5em
  minus 0.4em\relax Cambridge, UK: Cambridge University Press, 2004.

\bibitem{worldBank}
J.~Aron and I.~Elbadawi, ``{Foreign exchange auction markets in sub-Saharan
  Africa : dynamic models for auction exchange rates},'' \emph{The World Bank,
  Policy Research Working Paper Series}, January 1994.

\bibitem{cvAuctions}
J.~H. Kagel and D.~Levin, \emph{{Common Value Auctions and the Winner's
  Curse}}.\hskip 1em plus 0.5em minus 0.4em\relax Princeton, NJ: Princeton
  University Press, 2002.

\bibitem{entryCosts}
S.~Chakravarty and S.~Ghosh, ``{Studying the Effect of Sunk Costs on Bidding
  Behavior in Auctions},'' in \emph{{2007 Economic Science Association World
  Meetings}}, 2007.

\bibitem{bidPrepCosts}
W.~F. Samuelson, ``{Competitive bidding with entry costs},'' \emph{Economics
  Letters}, vol.~12, no. 1-2, pp. 53--57, March 1985.

\end{thebibliography}
\bibliographystyle{IEEEtran}

\appendix
\subsection{Formal characteristics of Maker auctions}
\label{section:characteristics}
The following section defines the characteristics of Maker auctions, within the framework of formal auction theory.

\textbf{Ascending}: Maker auctions are categorized as ascending, or English, auctions, in that the price of the reward (the collateral) starts low and continues to rise throughout the auction process \cite[p. 11]{auctionTheory.Klemperer}. The auctions do not, however, follow the process of Japanese auctions (a variant of English auctions), in which the auctioneer raises the price until all bidders drop out \cite[p. 187]{auctionTheory.Milgrom}. In the case of Maker auctions, individuals must submit their own bids to raise the price, and they are allowed to make ``jump bids'', which are bids that significantly increase the price above the minimum bid increments \cite[p. 11]{auctionTheory.Klemperer}.

\textbf{Second-price}: Maker auctions are equivalent to second-price auctions; if all bidders bid up to their reservation price (the highest price they are willing to pay), the winning bidder pays an amount equal to the reservation price of the second-highest bidder, adjusted for the minimum bid increment \cite[p. 10]{auctionTheory.Milgrom}. Maker auctions are not Vickrey auctions, however, as bids are not sealed. Therefore, bidders can learn about other bidders' behaviors, \textit{ex post}, by reviewing transaction data on the Ethereum blockchain (a method we use ourselves in Section~\ref{sec:optimize}).

The information to be gleaned about other bidders has two limitations. First, the only public information about each bidder is their address on the Ethereum blockchain. A bidder could easily use multiple addresses, which means that any analysis of bidding history by address would be unable to capture, with certainty, the full behavior of individual bidders. Second, the auctions do not have a formal drop-out mechanism, whereby bidders can formally signal they have ceased bidding. Bidders are able to submit a bid at any point in the auction, and they do not need to signal their entrance or withdrawal from an auction. Therefore, without knowing when other bidders have dropped out of a specific auction, a bidder is unable to collect information on the relative valuations of other bidders while an auction is ongoing.

\balance
\textbf{Single-unit}: Maker auctions are single-unit, as each auction sells a specified collateral amount, and each bidder must bid for the entirety of the collateral, without any adjustments to quantity \cite{worldBank}. There can be multiple single-unit auctions that run simultaneously, depending on the depth of liquidation volume.

\textbf{Interdependent-value}: Maker auctions have a common-value component to them, albeit with a departure from the conventional definition of common-value auctions. In the traditional model of common-value auctions, the item being auctioned has a single true value, but bidders do not know the true value \textit{ex ante} \cite[p. 2]{cvAuctions}. In Maker auctions, the collateral being auctioned has a true value, but bidders actually do know the true value of the collateral. For example, the value of ETH is defined by its price in US dollars. Each bidder has access to public information on the price of ETH, and would therefore know the true value of the collateral.

Maker auctions also have a private-value component, because of the participation costs required for bidders. These costs (which are defined in greater detail in Section~\ref{sec:costs}) must be incurred in order to participate in the auctions process, and therefore lower each bidder's valuation. The costs can be optimized, but the estimates required to formulate an optimization are such that each bidder will likely have a different optimization curve (see Section~\ref{sec:optimize} for further detail). Prior research has confirmed that bidders will adjust their bids to account for participation costs such as entrance fees or bid preparation costs \cite{entryCosts, bidPrepCosts}. The exact nature of the bid adjustment is theorized in the literature, but these theories do not translate specifically to Maker auctions. Participation costs in Maker auctions contain several nuances that are not accounted for in general models, such as transaction fees dependent on frequency of rebalancing, as well as the concept of slippage when exchanging currencies.

An additional private-value component comes from the potential usage of the collateral. For example, by virtue of superior knowledge or resources, a bidder may anticipate gaining additional profits after obtaining ETH, in excess of what an average bidder would expect to make. These private values are difficult to quantify, but are discussed as potential factors in auction valuations in Section~\ref{sec:optimize}.

In summary, Maker auctions have both common-value and private-value components. As a result, we categorize these auctions as interdependent-value auctions.

\end{document}